\documentclass[letterpaper]{article} 
\usepackage{aaai2026}  
\usepackage{times}  
\usepackage{helvet}  
\usepackage{courier}  
\usepackage[hyphens]{url}  
\usepackage{graphicx} 
\usepackage{natbib}  
\usepackage{caption} 
\usepackage{newfloat}
\usepackage{listings}
\DeclareCaptionStyle{ruled}{labelfont=normalfont,labelsep=colon,strut=off} 
\usepackage{float}
\floatstyle{ruled}
\newfloat{listing}{tb}{lst}{}
\floatname{listing}{Listing}
\usepackage{xspace}

\usepackage{algorithm}  
\usepackage{algorithmicx}  
\usepackage{listings}
\usepackage{algpseudocode}  
\usepackage{threeparttable} 

\newcommand{\ie}{\textit{i.e.,}\xspace}
\newcommand{\eg}{\textit{e.g.,}\xspace}

\newcommand{\first}{\textsf{(i)}\xspace}
\newcommand{\second}{\textsf{(ii)}\xspace}
\newcommand{\third}{\textsf{(iii)}\xspace}

\newcommand{\parab}[1]{\vspace{0.01in}\noindent{\bf #1} }
\usepackage{graphicx}
\usepackage{subcaption} %
\usepackage{amsmath}
\usepackage{amsfonts} 
\usepackage{amssymb} 
\usepackage{booktabs}
\usepackage{multirow}
\usepackage[table]{xcolor}

\newcommand{\sys}{QuFid\xspace} 

\title{Adaptive Fidelity Estimation for Quantum Programs with Graph-Guided Noise Awareness}

\author{
	Tingting Li\textsuperscript{\rm 1,}\textsuperscript{\rm 2},
    Ziming Zhao\textsuperscript{\rm 2,}\footnotemark[1],
	Jianwei Yin\textsuperscript{\rm 2,}\thanks{Corresponding authors.}
}
\affiliations{
	\textsuperscript{\rm 1}Shanghai Qi Zhi Institute, Shanghai, China \\
	\textsuperscript{\rm 2}Zhejiang University, Hangzhou, China \\
	\{litt2020, zhaoziming\}@zju.edu.cn, zjuyjw@cs.zju.edu.cn
}

\begin{document}

\maketitle

\begin{abstract}
Fidelity estimation is a critical yet resource-intensive step in testing quantum programs on noisy intermediate-scale quantum (NISQ) devices, where the required number of measurements is difficult to predefine due to hardware noise, device heterogeneity, and transpilation-induced circuit transformations.
We present \sys, an adaptive and noise-aware framework that determines measurement budgets online by leveraging circuit structure and runtime statistical feedback. \sys models a quantum program as a directed acyclic graph (DAG) and employs a control-flow-aware random walk to characterize noise propagation along gate dependencies. Backend-specific effects are captured via transpilation-induced structural deformation metrics, which are integrated into the random-walk formulation to induce a noise-propagation operator. Circuit complexity is then quantified through the spectral characteristics of this operator, providing a principled and lightweight basis for adaptive measurement planning.
Experiments on 18 quantum benchmarks executed on IBM Quantum backends show that \sys significantly reduces measurement cost compared to fixed-shot and learning-based baselines, while consistently maintaining acceptable fidelity bias.

\end{abstract}

\section{Introduction}\label{sec:intro}

Quantum computing exploits quantum-mechanical phenomena such as superposition and entanglement to enable advances in optimization~\cite{urbanek2020error,ayanzadeh2023enigma}, simulation~\cite{zhu2024quantum, xiang2024enhanced}, and material science~\cite{nielsen2001quantum}. Promising applications have already emerged in quantum simulation~\cite{preskill2018quantum}, quantum machine learning~\cite{li2024minerva, biamonte2017quantum}, and cryptography~\cite{shor1999polynomial, xu2024routing}.

However, in the current Noisy Intermediate-Scale Quantum (NISQ) era, quantum devices remain severely constrained by decoherence, gate errors, limited qubit counts, and the absence of full-scale error correction~\cite{preskill2018quantum,hu2023battle,liang2023hybrid,campbell2024series}. These noise sources degrade computational accuracy and make reliable state evaluation inherently challenging, hindering practical deployment and necessitating careful device characterization prior to execution. Benchmarking techniques such as randomized benchmarking (RB)~\cite{magesan2012characterizing} and cross-entropy benchmarking (XEB)~\cite{boixo2018characterizing} assess device performance through repeated circuit executions and statistical analysis~\cite{sun2014tracking}. In practice, however, the number of repetitions (shots) is typically chosen heuristically: insufficient shots lead to inaccurate fidelity estimates, while excessive shots waste scarce quantum resources.

Recent efforts have explored fidelity prediction and reliability estimation using pre-characterized noise models or learning-based approaches~\cite{tan2023quct,hanruiwang2022quantumnas}. While effective in controlled settings, these methods rely on prior knowledge of device-specific noise characteristics, which are often time-varying and difficult to obtain in real-world environments~\cite{li2025task}. As a result, there remains a pressing need for adaptive fidelity estimation strategies that can dynamically determine measurement budgets while accounting for both circuit structure and hardware-induced effects.

Through our investigation, we identify three fundamental challenges in determining the number of repetitions required for quantum benchmarking.
\first \textit{Noise heterogeneity.} Quantum noise, including both Markovian and non-Markovian components~\cite{breuer2002theory}, varies across devices and over time~\cite{kandala2019error}, making static or universal repetition rules unreliable.
\second \textit{Circuit-hardware discrepancy.} During transpilation, qubit mapping, circuit rewriting, and optimization~\cite{mckay2018qiskit} reshape the circuit's dependency structure and noise exposure~\cite{li2025empowering}. Existing approaches lack a unified way to quantify the resulting structural deformation and its impact on measurement reliability.
\third \textit{Precision-latency trade-off.} Fidelity evaluation must balance statistical accuracy (\eg for validation and verification) against rapid turnaround (\eg for calibration and testing), calling for adaptive and uncertainty-aware measurement strategies.

To address these challenges, we propose \sys, an adaptive and noise-aware fidelity measurement framework. \sys models a quantum program as a directed acyclic graph (DAG) and constructs a control-flow-aware random walk to characterize noise propagation along gate dependencies. Backend-specific effects introduced by transpilation are captured through \emph{structural deformation metrics}, which quantify changes in effective connectivity and dependency structure after mapping. These metrics are integrated into the random-walk formulation, inducing a noise-propagation operator whose spectral characteristics provide a principled and lightweight measure of circuit complexity. Based on this structural characterization and runtime statistical feedback, \sys adaptively allocates measurement budgets and applies confidence-driven early stopping to avoid redundant sampling while preserving estimation accuracy.

In summary, this paper makes the following contributions.
\begin{itemize}
    \item We identify fundamental limitations of existing fidelity estimation strategies and highlight the need for adaptive, structure-aware measurement planning under realistic noise conditions.
    \item We propose \sys, a novel framework that integrates control-flow-aware noise propagation, transpilation-induced structural deformation modeling, and spectral complexity analysis to guide adaptive fidelity measurement for quantum circuits.
    \item We evaluate \sys on 18 diverse quantum benchmarks executed on real IBM Quantum backends, demonstrating a significant reduction in measurement cost while maintaining high accuracy. Additional studies and extended analysis further validate its scalability and adaptability.
\end{itemize}

\section{Background and Related Work}\label{sec:related}

In this section, we review relevant background on quantum computing and fidelity measurement, survey existing approaches to measurement optimization, and motivate the need for adaptive fidelity estimation strategies.

\parab{Quantum Computing in the NISQ Era.}
Classical computers process information using deterministic bits and logic gates, whereas quantum computers operate on qubits that can exist in superposition,
\(\lvert \psi \rangle = \alpha \lvert 0 \rangle + \beta \lvert 1 \rangle\) with \(|\alpha|^2 + |\beta|^2 = 1\),
and exploit superposition, entanglement, and interference to explore multiple computational paths simultaneously~\cite{huang2021quantum}. 
In the Noisy Intermediate-Scale Quantum (NISQ) era~\cite{li2025fortuna}, quantum devices are constrained by limited qubit counts, high gate error rates, decoherence, and the absence of fault-tolerant error correction~\cite{preskill2018quantum}. 
As a result, quantum computations are inherently noisy and probabilistic, requiring repeated measurements and statistical analysis to obtain reliable results~\cite{preskill2018quantum}, making accurate state and fidelity evaluation both essential and resource-intensive.

\parab{Fidelity Measurement and Optimization.}
Fidelity measurement is a fundamental technique for validating and characterizing quantum devices and programs~\cite{gilyen2022improved,wang2022quantum,huang2020predicting}. 
Due to the high cost of repeated measurements, prior work has explored fidelity prediction and measurement optimization using machine learning~\cite{zhang2021direct,yu2022statistical,liu2020reliability}, variational algorithms~\cite{cerezo2020variational,chen2021variational,tan2021variational}, and classical-shadow-based techniques~\cite{huang2020predicting}. 
For example, QuEst~\cite{wang2022quest} applies graph transformers to predict circuit fidelity, while QuCT~\cite{tan2023quct} employs a unitary-to-vector representation. 
While effective in certain scenarios, these approaches typically rely on pre-characterized noise models or historical training data and focus on predicting fidelity values rather than determining the minimal number of measurements required to achieve a desired accuracy, limiting their applicability under dynamic and heterogeneous hardware conditions.

\parab{Random Walk Models and Structural Analysis.}
Random walks model stochastic traversal over graph structures based on transition probabilities~\cite{tong2006fast} and are widely used to capture both local and global structural properties. 
In quantum computing~\cite{li2024qust}, random-walk-based techniques have been applied to analyze state evolution, circuit structure, and dependency patterns~\cite{qiang2016efficient}, as well as to support noise-aware synthesis and optimization~\cite{bergholm2018pennylane}. 
More broadly, random walks are extensively studied in classical graph domains for information diffusion, graph property estimation, and Monte Carlo optimization~\cite{tong2006fast,li2015random,craswell2007random}. 
Most existing applications use random walks as heuristic tools for structural characterization or sampling, whereas our work leverages them to model noise propagation along control-flow dependencies, enabling an operator-level abstraction that connects circuit structure, hardware-induced deformation, and measurement uncertainty.

\parab{Motivation and Positioning.}
Determining an appropriate number of measurements remains a central challenge in quantum benchmarking and testing. This challenge arises from heterogeneous and time-varying noise~\cite{breuer2002theory,kandala2019error}, discrepancies between logical circuits and their transpiled implementations~\cite{mckay2018qiskit}, and the trade-off between statistical accuracy and measurement latency. 
In practice, shot counts are often chosen heuristically, leading to unreliable estimates or unnecessary resource consumption~\cite{wang2022quest}. 
Existing approaches~\cite{tan2023quct,hanruiwang2022quantumnas} largely assume static noise characteristics or offline training, limiting their effectiveness in realistic NISQ environments. 
Motivated by these limitations, we propose an adaptive fidelity measurement framework that integrates structural circuit analysis with runtime statistical feedback. By modeling transpilation-induced structural deformation and noise propagation within a unified graph-based abstraction, our approach enables principled, backend-aware measurement budgeting that adapts to both circuit structure and observed convergence behavior.

\section{Methodology}\label{sec:method}

\begin{algorithm}[t]
	\caption{Adaptive Fidelity Measurement Pipeline}
	\label{alg:adaptive}
	\begin{algorithmic}[1]
	\Require Quantum circuit $Q$, target error bound $\delta$, confidence level $\alpha$
	\Ensure Estimated fidelity $\hat{F}$
	\State $\mathcal{G} \leftarrow$ Convert $Q$ into a DAG
	\State Construct graph-structural deformation metrics on $\mathcal{G}$
	\State Build control-flow-aware noise-propagation operator $P$
	\State Compute spectral complexity $C(\mathcal{G})$ from $P$
	\State $\mathcal{D} \leftarrow$ Depth of transpiled circuit
	\State $\mathcal{P} \leftarrow C(\mathcal{G}) \cdot \log(\mathcal{D})$ \Comment{Adaptive batch size}
	\State Initialize measurement set $T \leftarrow \emptyset$
	\Repeat
		\State Execute $\mathcal{P}$ measurements $\rightarrow \{T_i\}$
		\State $T \leftarrow T \cup \{T_i\}$
		\State $\hat{F} \leftarrow \mathrm{mean}(T)$
		\State $\sigma \leftarrow \mathrm{std}(T)$
		\State $CI \leftarrow z_{\alpha} \cdot \frac{\sigma}{\sqrt{|T|}}$
		\If{$CI \leq \delta$}
			\State \Return $\hat{F}$ \Comment{Confidence-driven early stopping}
		\EndIf
	\Until{$|T| \geq \mathcal{P}_{\max}$}
	\State \Return $\hat{F}$
	\end{algorithmic}
\end{algorithm}

\begin{figure}[t]
    \centering
    \includegraphics[width=0.46\textwidth]{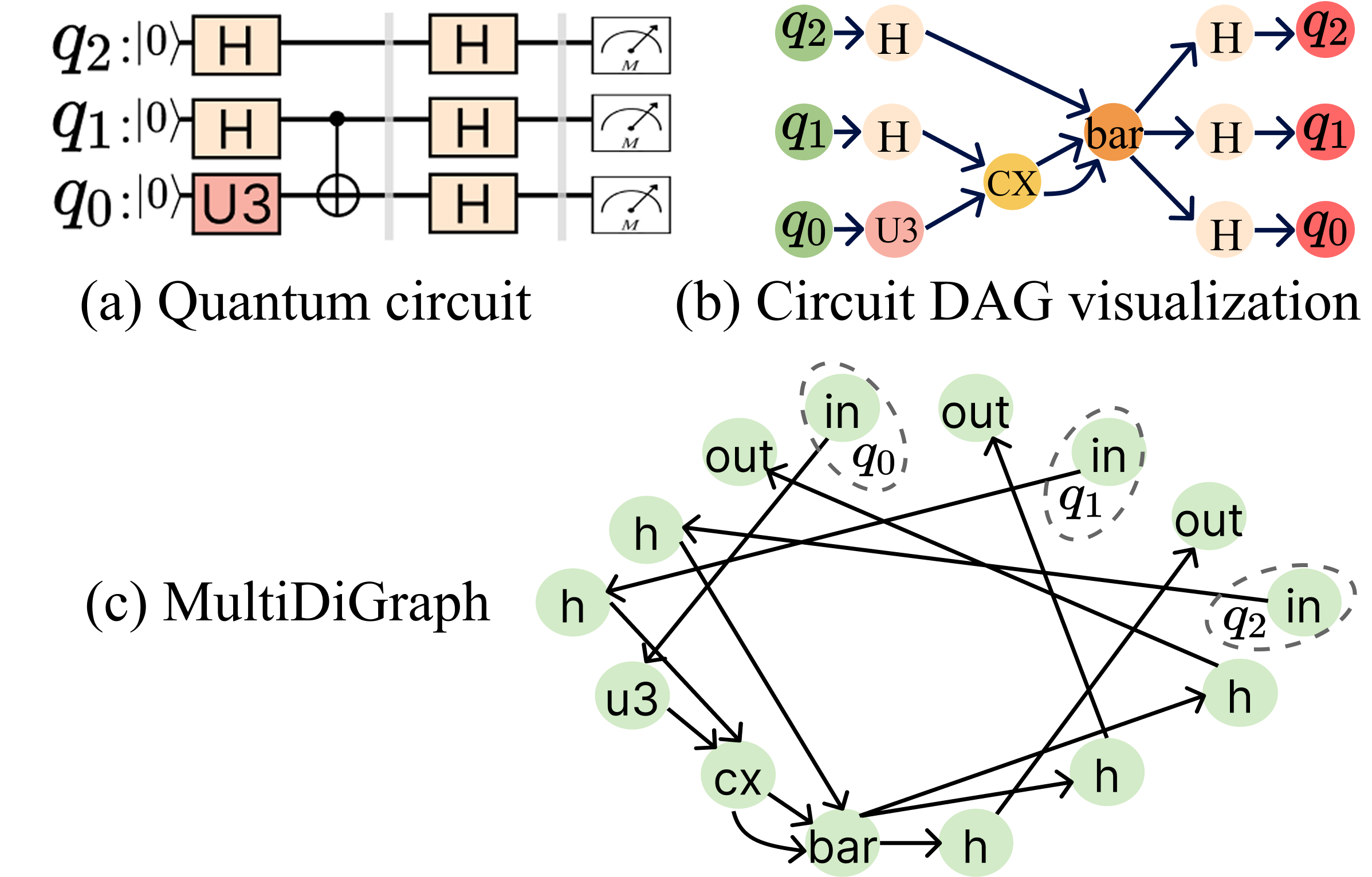}
    \caption{The DAG conversion and multi-directed graph of the BV algorithm quantum circuit.}
    \label{fig:dag}
\end{figure}

We model fidelity estimation as a structure-aware analysis problem that connects circuit dependencies, noise propagation, and statistical convergence.
Rather than using random walks as a direct complexity estimator~\cite{tong2006fast}, we employ a control-flow-aware random walk as an abstraction for how noise diffuses through gate dependencies in a quantum circuit~\cite{masuda2017random, fontana2025classical}.
This formulation induces a noise-propagation operator whose structural properties reflect the circuit's intrinsic complexity after transpilation.
Circuit complexity is then quantified through the spectral characteristics of this operator, providing a principled and lightweight basis for determining adaptive measurement iterations~\cite{coifman2006diffusion}.
The overall adaptive measurement pipeline is summarized in Algorithm~\ref{alg:adaptive}.

\parab{Quantum Circuit Graph Construction.}
We represent a quantum program as a directed acyclic graph (DAG) $\mathcal{G}=(V,E)$, where each node $v_i \in V$ corresponds to a quantum gate and each directed edge $(v_i, v_j) \in E$ encodes a non-commutative control-flow dependency induced by shared qubits.
A directed edge indicates that gate $v_j$ cannot be executed before $v_i$ without altering circuit semantics, thereby explicitly capturing execution-order constraints and inter-gate interactions.
This DAG representation provides a unified abstraction for both logical circuit structure and dependency-induced error propagation.
As shown in Figure~\ref{fig:dag}, we first convert the Bernstein-Vazirani circuit into a DAG~\cite{qiskit} and then into a MultiDiGraph to capture topological connections. Each gate corresponds to a node, and qubits define edges. Nodes store gate type, target qubits, and parameters, while edges encode non-commutative relationships~\cite{smith2023clifford}.

\parab{Structural Deformation Metrics.}
The transpilation transforms a logical circuit into a hardware-executable form by reordering gates, inserting auxiliary operations, and remapping qubits.
These transformations expose and reshape control-flow dependencies, often introducing additional interactions that affect noise accumulation and propagation.
Let $\mathcal{G}_0=(V_0,E_0)$ and $\mathcal{G}_t=(V_t,E_t)$ denote the circuit graphs before and after transpilation, respectively.
Transpilation induces \emph{structural deformation} by introducing auxiliary gates, modifying dependency paths, and altering connectivity patterns.
We quantify such deformation using graph-level metrics defined as relative structural changes:
\begin{equation}
\Delta_{\text{struct}}(\mathcal{G}_0,\mathcal{G}_t)
= \{\Delta_{\deg}, \Delta_{\text{path}}, \Delta_{\text{conn}}\},
\end{equation}
where $\Delta_{\deg}$ measures shifts in node degree distribution, $\Delta_{\text{path}}$ captures the expansion of critical paths and long dependency chains, and $\Delta_{\text{conn}}$ reflects effective connectivity inflation.
These metrics summarize how transpilation reshapes the circuit's structural footprint in a backend-aware yet noise-model-agnostic manner.

\begin{figure}[t]
	\centering
	\includegraphics[width=0.46\textwidth]{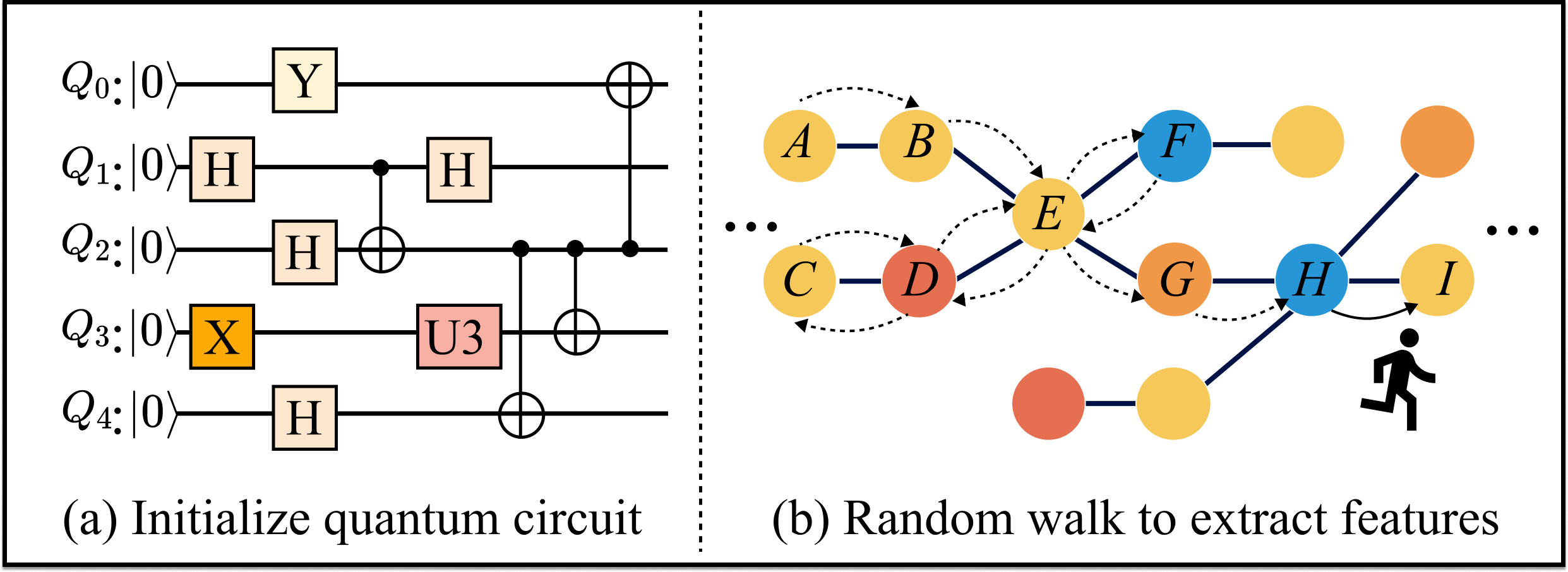}  
	\caption{Illustrative explanation of circuit analysis.} 
	\label{fig:overview}
\end{figure}

\parab{Control-Flow-Aware Noise Propagation Model.}
Based on the deformed circuit graph $\mathcal{G}_t$, we model noise propagation using a control-flow-aware random walk (Figure~\ref{fig:overview}) that captures how local gate errors accumulate and spread along dependency chains during circuit execution.
Let $A \in \mathbb{R}^{n\times n}$ denote a weighted adjacency matrix incorporating structural deformation, where $n = |V_t|$.
Each weight $A_{ij}$ reflects the strength of the control-flow dependency from gate $v_i$ to $v_j$, modulated by deformation-induced factors such as dependency expansion and effective connectivity inflation.
We define the corresponding noise-propagation operator
\begin{equation}
P = D^{-1}A, \quad D = \mathrm{diag}(d_1,\ldots,d_n), \; d_i = \sum_j A_{ij},
\end{equation}
which normalizes outgoing dependencies to form a row-stochastic transition matrix.
Each entry $P_{ij}$ represents the conditional probability that noise originating at gate $v_i$ influences gate $v_j$ through valid control-flow paths.
Repeated application of $P$ thus models a Markovian diffusion process over the circuit's dependency structure, providing an abstraction of cumulative noise propagation across execution steps.
Unlike uniform random walks, the deformation-aware weighting in $A$ biases propagation toward structurally critical paths, such as long dependency chains or high fan-in regions.
As a result, the induced operator $P$ reflects not only logical control flow but also backend-induced structural effects that amplify or attenuate noise transmission, making it a suitable foundation for structure-aware fidelity analysis.

\parab{Spectral Complexity Estimation.}
We quantify circuit complexity through the spectral characteristics of the noise-propagation operator $P$, which provide an intrinsic and time-independent description of how noise diffuses through circuit dependencies.
Let $\{\lambda_i\}_{i=1}^{n}$ denote the eigenvalues of $P$, ordered by decreasing magnitude.
Unlike procedural convergence metrics that depend on iterative dynamics, spectral components characterize global structural properties of the underlying graph and capture the strength, persistence, and reach of dependency-driven noise diffusion.
In particular, dominant eigenvalues correspond to slowly decaying propagation modes, indicating long-range coupling and strong interdependence among gates.
Such modes imply that local noise can influence a larger portion of the circuit and persist across execution steps, increasing the uncertainty of fidelity estimation.
We therefore define the circuit complexity as
\begin{equation}
C(\mathcal{G}) = \sum_{i=1}^{k} |\lambda_i|,
\end{equation}
where $k \ll n$ retains only the dominant spectral modes that contribute most significantly to noise propagation.
This truncated spectral aggregation yields a compact yet expressive complexity measure that is robust to graph size and minor structural variations.

Intuitively, a larger spectral mass indicates stronger long-range dependencies and slower attenuation of noise influence, suggesting that more measurements are required for empirical fidelity estimates to converge.
Conversely, circuits with rapidly decaying spectra exhibit limited noise diffusion and require fewer samples to achieve stable estimates.
By grounding complexity estimation in spectral properties, our approach provides a principled and backend-aware basis for adaptive measurement planning.

\begin{figure}[t]
	\centering
	\includegraphics[width=1\linewidth]{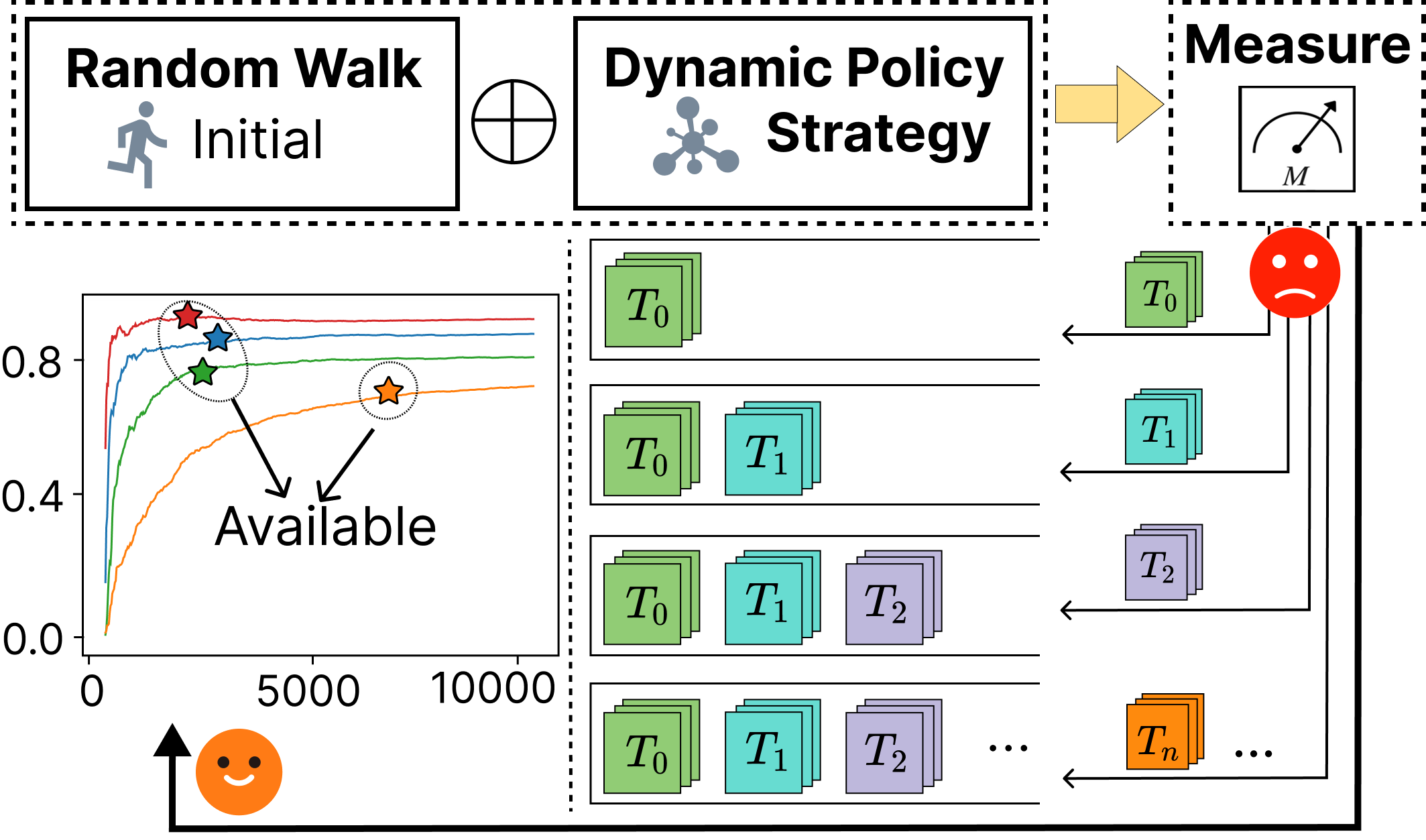}  
	\caption{Adaptive measurement planning procedure.}
	\label{fig:iter}
\end{figure}

\parab{Adaptive Measurement Planning.}
Fidelity estimation is formulated as an adaptive testing process guided by structural complexity.
Given the estimated spectral complexity $C(\mathcal{G})$ and the transpiled circuit depth $\mathcal{D}$, we determine the batch size as
\begin{equation}
\mathcal{P} = C(\mathcal{G}) \cdot \log(\mathcal{D}),
\end{equation}
which assigns larger measurement budgets to circuits exhibiting stronger structural deformation and deeper dependency chains.
This logarithmic scaling with depth prevents over-allocation for deep but weakly coupled circuits, while ensuring sufficient sampling for circuits with pronounced long-range dependencies.
Measurements are executed iteratively, and fidelity statistics are updated after each batch, allowing sampling effort to be dynamically adjusted based on structural characteristics and convergence behavior.
Figure~\ref{fig:iter} illustrates this adaptive batching process.
At each iteration, a batch of $\mathcal{P}$ measurements is executed and aggregated with previous results, and the process continues until the uncertainty-aware stopping criterion is satisfied.

\parab{Uncertainty-Aware Termination.}
To ensure statistical reliability, we employ a confidence-interval-based stopping criterion that explicitly controls estimation uncertainty.
Let $\hat{F}$ be the empirical fidelity estimate and $\sigma$ its standard deviation.
Sampling terminates once
\begin{equation}
z_{\alpha} \cdot \frac{\sigma}{\sqrt{|T|}} \leq \delta,
\end{equation}
ensuring that the estimation error does not exceed $\delta$ with confidence level $(1-\alpha)$.
This uncertainty-aware policy complements the structure-guided batching strategy by halting measurements as soon as statistical stability is achieved, thereby avoiding redundant sampling while preserving rigorous accuracy guarantees.

Overall, our methodology formulates fidelity estimation as a structure-aware and uncertainty-driven test planning problem.
By integrating structural deformation analysis, control-flow-aware noise propagation, spectral complexity estimation, and adaptive stopping, \sys provides an efficient fidelity oracle for noisy quantum programs.

\section{Evaluation}

\begin{table*}[h]
    \centering
    \resizebox{\linewidth}{!}{
    \begin{tabular}{l|l||l|l||l|l}
    \toprule
    \textbf{Notation} & \textbf{Quantum Circuit} 
    & \textbf{Notation} & \textbf{Quantum Circuit}
    & \textbf{Notation} & \textbf{Quantum Circuit} \\ \hline \hline
    BV & Bernstein-Vazirani
    & Clifford & Random Clifford circuit
    & Ising & Linear Ising model \\
    QAOA & Quantum Approx. Optimization
    & VQE & Variational Quantum Eigensolver
    & QFT & Quantum Fourier Transform \\
    QKNN & Quantum k-NN
    & QNN & Quantum Neural Network
    & QPE & Quantum Phase Estimation \\
    QSVM & Quantum Support Vector Machine
    & QuGAN & Quantum GAN
    & RB & Randomized Benchmarking \\
    Amplitude & Amplitude Estimation
    & Shor & Shor's factoring algorithm
    & Simon & Simon's algorithm \\
    SU2 & SU(2) circuits
    & VQC & Variational Quantum Circuit
    & XEB & Cross-Entropy Benchmarking \\
    \bottomrule
    \end{tabular}}
    \caption{Quantum circuit benchmarks used in our experiments.}
    \label{tab:bench}
\end{table*}

\begin{table*}[t]
    \centering
    \resizebox{1.0\textwidth}{!}{
    \begin{tabular}{c|c|rrrr rrrrrr rrrrr}
        \toprule
        \multicolumn{2}{c}{\textbf{Iter.}} & \multicolumn{4}{c}{BV} & \multicolumn{1}{c}{} & \multicolumn{4}{c}{QKNN}  & \multicolumn{1}{c}{} & \multicolumn{4}{c}{QPE}\\ \cline{3-6} \cline{8-11} \cline{13-16}
        \multicolumn{2}{c}{\textbf{Times}} & \multicolumn{1}{c}{4 qubits} & \multicolumn{1}{c}{6 qubits} & \multicolumn{1}{c}{8 qubits} & \multicolumn{1}{c}{10 qubits} & \multicolumn{1}{c}{} & \multicolumn{1}{c}{4 qubits} & \multicolumn{1}{c}{6 qubits} & \multicolumn{1}{c}{8 qubits} & \multicolumn{1}{c}{10 qubits} & \multicolumn{1}{c}{} & \multicolumn{1}{c}{4 qubits} & \multicolumn{1}{c}{6 qubits} & \multicolumn{1}{c}{8 qubits} & \multicolumn{1}{c}{10 qubits} \\ \hline \hline
\multirow{15}{*}{\rotatebox{90}{\textbf{Vanilla iteration}}} &     \textbf{50}  & 0.0062 & 0.0308 & 0.1416 & 0.0247 &  & 0.3889 & 0.5318 & 0.9321 & 0.8609 &  & \cellcolor{gray!50} 0.0516 & \cellcolor{gray!50} 0.0728 & 0.0648 & 0.0700  \\
 &  \textbf{550} & \cellcolor{gray!50} 0.0150 & \cellcolor{gray!50} 0.0077 & \cellcolor{gray!50} 0.0358 & \cellcolor{gray!50} 0.0035  & & 0.0255 & 0.0429 & 0.1646 & 0.2574 &  & \cellcolor{gray!50} 0.0080 & \cellcolor{gray!50} 0.0022 & 0.0179 & 0.0162  \\
 &    \textbf{1050}& \cellcolor{gray!50} 0.0075 & \cellcolor{gray!50} 0.0014 & \cellcolor{gray!50} 0.0083 & \cellcolor{gray!50} 0.0028  & & \cellcolor{gray!50} 0.0135 & 0.0260 & 0.1084 & 0.1290 &  & 0.0092 & 0.0047 & \cellcolor{gray!50} 0.0156 & 0.0063  \\
 &    \textbf{1550}& 0.0056 & 0.0025 & 0.0008 & 0.0004  & & \cellcolor{gray!50} 0.0100 & 0.0151 & 0.0662 & 0.0891 &  & 0.0070 & 0.0044 & \cellcolor{gray!50} 0.0010 & \cellcolor{gray!50} 0.0141  \\
 &    \textbf{2050}& 0.0012 & 0.0051 & 0.0079 & 0.0020  & & 0.0069 & 0.0117 & 0.0392 & 0.0634 &  & 0.0048 & 0.0024 & 0.0010 & \cellcolor{gray!50} 0.0051  \\
 &    \textbf{2550}& 0.0001 & 0.0078 & 0.0060 & 0.0027  & & 0.0002 & \cellcolor{gray!50} 0.0110 & 0.0264 & 0.0506 &  & 0.0040 & 0.0042 & 0.0050 & 0.0033  \\
 &    \textbf{3050}& 0.0026 & 0.0056 & 0.0018 & 0.0021  & & 0.0006 & \cellcolor{gray!50} 0.0064 & 0.0211 & 0.0441 &  & 0.0030 & 0.0036 & 0.0041 & 0.0034  \\
 &    \textbf{3550}& 0.0014 & 0.0049 & 0.0010 & 0.0048  & & 0.0013 & 0.0045 & 0.0180 & 0.0326 &  & 0.0029 & 0.0012 & 0.0051 & 0.0018  \\
 &    \textbf{4050}& 0.0005 & 0.0042 & 0.0006 & 0.0047  & & 0.0009 & 0.0028 & 0.0153 & 0.0273 &  & 0.0001 & 0.0004 & 0.0039 & 0.0001  \\
 &    \textbf{4550}& 0.0016 & 0.0049 & 0.0009 & 0.0032  & & 0.0018 & 0.0034 & 0.0138 & 0.0226 &  & 0.0002 & 0.0018 & 0.0027 & 0.0058  \\
 &    \textbf{5050}& 0.0008 & 0.0035 & 0.0024 & 0.0064  & & 0.0013 & 0.0036 & \cellcolor{gray!50} 0.0109 & 0.0189 &  & 0.0018 & 0.0029 & 0.0007 & 0.0015  \\
 &    \textbf{6050}& 0.0003 & 0.0014 & 0.0018 & 0.0027  & & 0.0008 & 0.0020 & \cellcolor{gray!50} 0.0073 & \cellcolor{gray!50} 0.0161 &  & 0.0006 & 0.0005 & 0.0002 & 0.0029  \\
 &    \textbf{7050}& 0.0009 & 0.0000 & 0.0026 & 0.0014  & & 0.0006 & 0.0019 & 0.0052 & \cellcolor{gray!50} 0.0081 &  & 0.0004 & 0.0016 & 0.0019 & 0.0019  \\
 &    \textbf{8050}& 0.0013 & 0.0016 & 0.0020 & 0.0011  & & 0.0007 & 0.0002 & 0.0047 & 0.0045 &  & 0.0024 & 0.0026 & 0.0014 & 0.0022  \\
 &    \textbf{9050}& 0.0005 & 0.0005 & 0.0017 & 0.0002  & & 0.0004 & 0.0005 & 0.0022 & 0.0011 &  & 0.0008 & 0.0021 & 0.0002 & 0.0004  \\ \hline \hline
 \rowcolor{gray!30}   \multicolumn{2}{c}{\textbf{Our bias}} & 0.0071 & 0.0017 & 0.0084 & 0.0033 &  & 0.0076 & 0.0086 & 0.0092 & 0.0082 &  & 0.0039 & 0.0022 & 0.0008 & 0.0064  \\ \hline
 \rowcolor{gray!30}   \multicolumn{2}{c}{\textbf{Our iter.}} & 592    & 792   & 860   & 981  &   & 1192   & 2827   & 5584   & 6992 &   & 173    & 954    & 1570   & 1906   \\
    \bottomrule
    \end{tabular}}
    \caption{Comparison of the vanilla iterative processes and the proposed scheme for 4 qubit sizes in 3 circuits, \ie `BV', `QKNN', and `QPE', with setting the fidelity bias to less than 0.01. `Our iter' and `Our bias' correspond to required iterations  and achieved fidelity bias. `Iter. Times' refers to the number of vanilla iterations.}
    \label{tab:base}
\end{table*}

This section describes the experimental design and infrastructure used to evaluate \sys.
We first outline the evaluation objectives, followed by the experimental environment, benchmark circuits, and implementation details.
Our evaluation is designed to assess \sys from multiple complementary perspectives.
Specifically, we examine whether \sys can reduce the number of fidelity measurements required to achieve a fixed precision target, and how its adaptive behavior trades off estimation accuracy against measurement cost under varying test budgets.
We further analyze the quality of measurement allocation, including test effectiveness, scalability, and adaptability.

\parab{Experimental Platform and Settings.}
All experiments are conducted on IBM Quantum hardware platforms~\cite{ibmq} via the Qiskit SDK~\cite{qiskit}. Specifically, we use three IBM backends: \texttt{Sherbrooke}, \texttt{Kyiv}, and \texttt{Brisbane}, each offering up to 127 qubits. The default fidelity measurement for baselines is set to \texttt{shots=10000}, with sampling intervals of \texttt{step=20} for iteration comparisons.
We set the confidence level $\alpha = 0.05$ and the default fidelity bias threshold $\delta = 0.01$. 

\begin{figure}[t]
    \centering
    \includegraphics[width=1\linewidth]{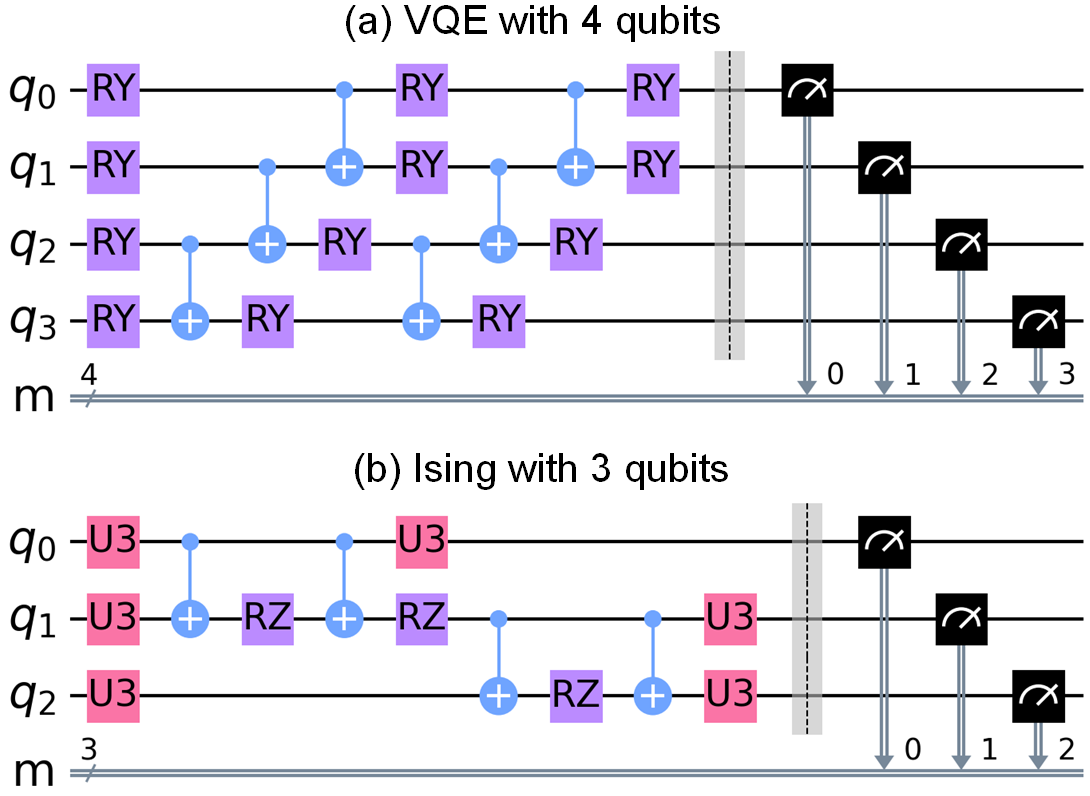}
    \caption{Some instances of quantum circuits.}
    \label{fig:circuit}
\end{figure}

\parab{Benchmark Circuits.}
We evaluate \sys on 18 representative quantum algorithms~\cite{zhang2023oneq,li2019tackling,li2020towards,li2020eliminating} widely used in software verification, optimization, and simulation. 
As summarized in Table~\ref{tab:bench}, these include classical algorithms (BV, Simon, QPE, QFT, Shor, Clifford), variational algorithms (QAOA, VQE, VQC, SU2), machine learning models (QKNN, QNN, QSVM, QuGAN), and benchmarking circuits (RB, XEB, and Amplitude Estimation).
Each circuit is evaluated under qubit sizes $\{4, 6, 8, 10\}$ and Figure~\ref{fig:circuit} illustrates example circuits.

\begin{figure*}[t]
	\centering
    \includegraphics[width=1\linewidth]{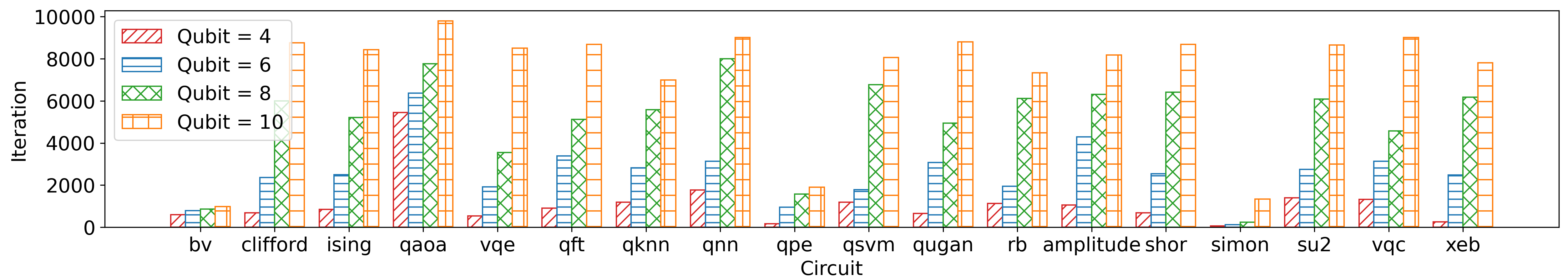}
	\caption{The number of iterations required for 18 circuits with 4 qubit sizes sets the fidelity bias to less than 0.01.}
	\label{fig:zhu}
\end{figure*}

\begin{figure*}[t]
	\centering
    \includegraphics[width=1\linewidth]{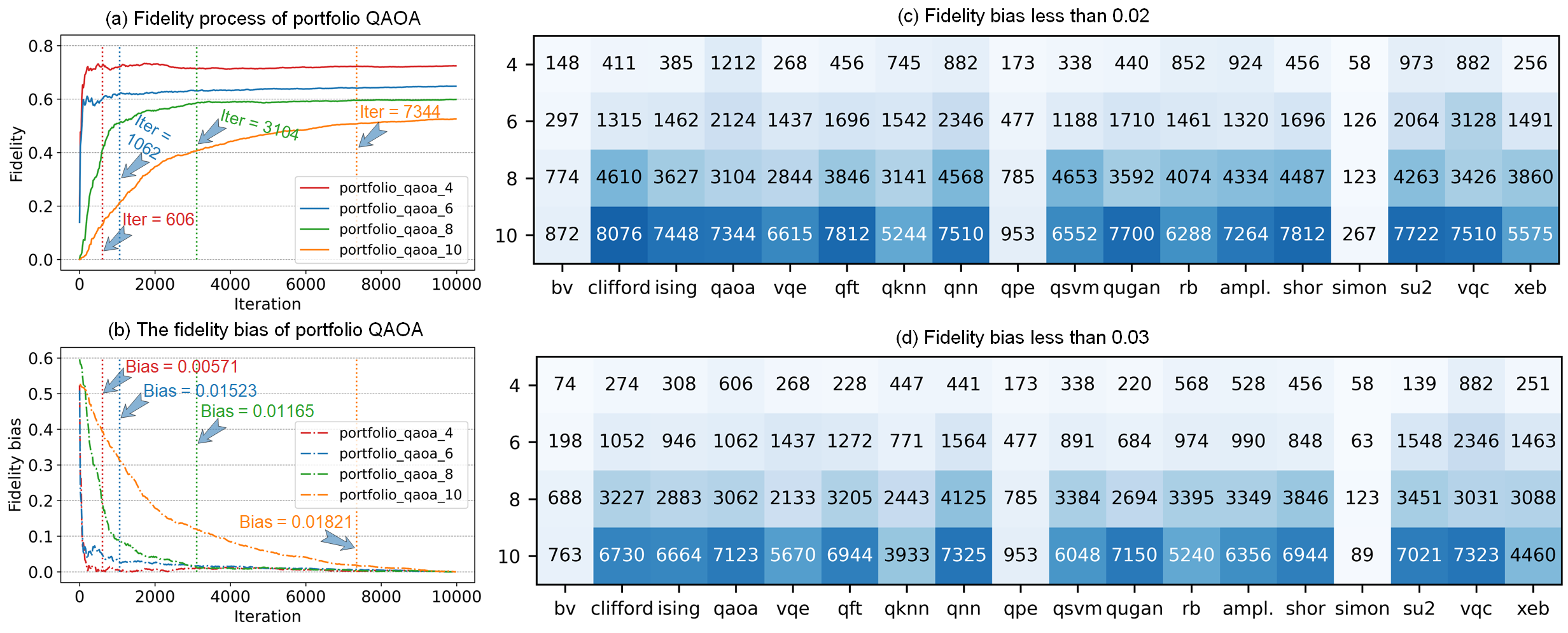}
	\caption{Case study and evaluation with different measurement biases.}
	\label{fig:f6}
\end{figure*}

\parab{Efficiency Evaluations.}
We evaluate 18 benchmarks to estimate the number of measurements required to achieve a fidelity deviation below $0.01$, as shown in Figure~\ref{fig:zhu}. 
The results show that required iterations generally increase with qubit count. 
Most algorithms, except BV, QPE, and Simon, need around 8000 measurements for 10-qubit circuits due to higher structural complexity, whereas 4-qubit circuits typically converge within a few hundred iterations. 
This variation indicates that using a fixed shot count can either waste resources or reduce accuracy, highlighting the need for adaptive measurement. 
Across all settings, \sys reduces measurement shots compared to the fixed-shot baseline.

\parab{Fidelity Measurement Trends.}
To further analyze fidelity evolution with increasing measurements, Table~\ref{tab:base} reports results for three representative algorithms, Bernstein-Vazirani (BV), Quantum k-Nearest Neighbor (QKNN), and Quantum Phase Estimation (QPE), on circuits with 4-10 qubits. 
Experiments are run on three IBM devices, and results are averaged. 
Absolute fidelity deviation is measured using the Hellinger distance between the estimated distribution and the ``true'' distribution obtained from 10,000 shots. 
Vanilla iteration uses a fixed sampling interval of 20 up to 10,000 iterations, with our evaluation sampled every 500 iterations. 
Gray shading denotes the estimated iteration bounds. 
All deviations are within 0.01.
As shown in Table~\ref{tab:base}, BV and QPE require relatively stable measurement counts as qubit size grows, while QKNN exhibits a steep increase: about 1,550, 2,550, 5,050, and 7,050 measurements for 4, 6, 8, and 10 qubits, respectively, reflecting higher structural complexity. 
In all cases, \sys maintains an absolute fidelity deviation below 0.01 while using significantly fewer measurements than baseline approaches, demonstrating robustness and efficiency for resource-constrained quantum experiments.

\parab{Case Study.}
We use the portfolio Quantum Approximate Optimization Algorithm (QAOA) as a case study to visualize how \sys adapts measurement counts and fidelity bias as iterations increase. 
Experiments are conducted on 4, 6, 8, and 10 qubits, with vertical dashed lines in Figures~\ref{fig:f6}(a)-(b) indicating the iteration points selected by \sys.
\textit{Measurement iterations.} 
As shown in Figure~\ref{fig:f6}(a), larger qubit sizes require more iterations to stabilize fidelity. 
For all qubit settings, fidelity increases initially, fluctuates briefly, and then converges. 
Stabilization occurs at approximately 606, 1062, 3104, and 7344 iterations for 4-, 6-, 8-, and 10-qubit circuits, respectively, consistent with the estimates made by \sys.
\textit{Fidelity bias.} 
Figure~\ref{fig:f6}(b) shows that fidelity bias decreases sharply at first, fluctuates, and then converges as iterations grow. 
Smaller qubit circuits converge faster: the 4-qubit circuit reaches a bias of 0.00571 in 606 iterations, while the 10-qubit circuit reaches 0.01821 after 7344 iterations. 
Overall, \sys achieves stable fidelity with fewer tests than fixed-shot methods, balancing accuracy and resource usage.

\parab{Different Measurement Biases.}
To study how bias tolerance affects measurement cost, we analyze the number of required measurements under different fidelity bias thresholds.
Figures~\ref{fig:f6}(c)-(d) report the measurement counts when the fidelity bias is constrained to be below $0.02$ and $0.03$, respectively.
Across all benchmarks and circuit sizes, relaxing the bias tolerance consistently reduces the required number of measurements, often by several thousand shots.
The reduction becomes more pronounced for deeper and structurally complex circuits.
For example, for the Ising benchmark with 10 qubits, the required measurement count decreases from 7448 under a bias threshold of $0.02$ to 6664 when the threshold is relaxed to $0.03$.
Similar trends are observed for Clifford, QFT, and QKNN circuits, indicating a clear trade-off between estimation accuracy and measurement cost.
These results confirm that \sys can effectively exploit looser accuracy requirements to substantially reduce testing overhead, while maintaining controlled fidelity bias.

\begin{figure}[t]
    \centering
    \includegraphics[width=\linewidth]{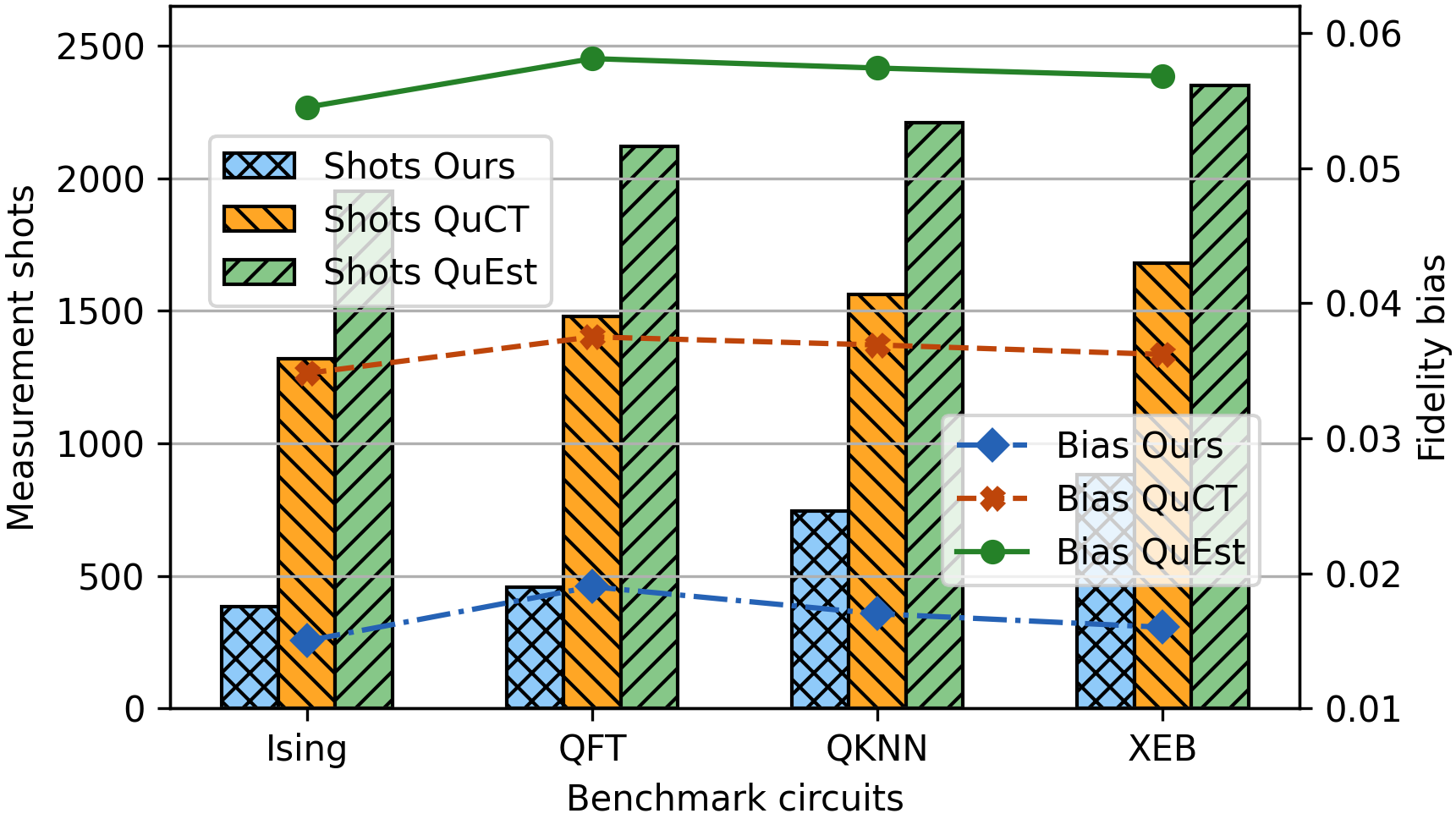}
    \caption{Comparison of measurement shots and bias.}
    \label{fig:rq5}
\end{figure}

\parab{Comparison with Existing Methods.}
We compare \sys with representative learning-based fidelity predictors, QuCT~\cite{tan2023quct} and QuEst~\cite{wang2022quest}.
Unlike QuCT and QuEst, which predict fidelity values from circuit features, \sys adaptively determines the minimal number of measurements needed to achieve reliable fidelity estimation. 
Figure~\ref{fig:rq5} reports both the measurement shots consumed (bar plots) and the resulting fidelity bias (line plots) across representative benchmark circuits.
Across all cases, \sys consistently requires substantially fewer measurement shots while achieving lower or comparable fidelity bias.
In contrast, QuCT incurs higher measurement cost due to its reliance on fine-grained feature modeling and validation, while QuEst exhibits the largest shot consumption and higher bias, reflecting the overhead and generalization limitations of graph-based prediction models.
These results highlight the advantage of \sys in jointly optimizing measurement efficiency and estimation accuracy under realistic NISQ settings.

\section{Discussion and Future Work}

This work reframes fidelity estimation for quantum programs as a structure-aware and uncertainty-driven test planning problem.
Rather than treating the number of measurements as a fixed hyperparameter or relying on pre-trained predictors, \sys dynamically allocates measurement budgets by jointly considering circuit structure, transpilation-induced deformation, and runtime statistical feedback.
The experimental results demonstrate that this perspective is both effective and practical under realistic NISQ conditions.

\parab{About Circuit Representation.}
A key insight of \sys is that circuit structure, especially after transpilation, plays a central role in determining how noise accumulates and propagates.
Our results show that circuits with similar depth or qubit counts can exhibit markedly different convergence behaviors due to differences in dependency patterns and effective connectivity.
By modeling these effects through a control-flow-aware noise-propagation operator and spectral complexity, \sys captures global structural characteristics that are missed by depth-based or gate-count-based heuristics.
This explains why \sys consistently adapts measurement budgets more accurately than fixed-shot strategies across diverse circuit families.

\parab{Interpretability and Practicality.}
Compared to learning-based approaches such as QuCT and QuEst, \sys does not require offline training, historical execution traces, or explicit noise models.
This makes \sys particularly well-suited for cold-start scenarios and time-varying hardware environments, where prior knowledge may be unavailable or outdated.
While learning-based methods aim to predict fidelity values directly, \sys focuses on determining the minimal number of measurements needed to achieve a desired accuracy, resulting in lower effective measurement cost and stronger robustness guarantees.
Each component of the framework, structural deformation metrics, spectral complexity, and confidence-driven stopping, has a clear physical or statistical meaning.
This transparency makes the system easier to reason about and debug, and allows practitioners to adjust accuracy targets or confidence levels in practice.

\parab{Limitations and Future Directions.}
Despite its effectiveness, \sys has several limitations that suggest directions for future work.
First, the current formulation assumes a Markovian abstraction of noise propagation, which may not fully capture long-range temporal correlations in certain devices.
Second, while spectral complexity provides a compact global measure, incorporating localized or qubit-specific deformation metrics may further improve adaptivity.
Finally, extending \sys to support other frameworks, \eg MindSpore Quantum~\cite{xu2024mindspore}, and multi-objective optimization, such as jointly optimizing fidelity, latency, and energy consumption, remains promising directions.

\section{Conclusion}\label{sec:conclusion}

In this paper, we present \sys, an adaptive and noise-aware framework that determines measurement budgets online by integrating structural circuit analysis with runtime statistical guarantees.
By modeling quantum programs as directed acyclic graphs and capturing transpilation-induced deformation through a control-flow-aware noise-propagation model, \sys quantifies circuit complexity via spectral characteristics and uses this information to guide adaptive measurement planning.
Extensive experiments on 18 representative quantum benchmarks executed on real IBM Quantum backends demonstrate that \sys substantially reduces measurement cost while maintaining strict fidelity accuracy constraints.
Compared with fixed-shot strategies and state-of-the-art learning-based predictors, \sys achieves a superior balance between accuracy and resource efficiency.
We believe that \sys represents a step toward principled, interpretable, and resource-efficient testing methodologies for quantum programs.

\section{Acknowledgments}
This research was supported by the Shanghai Qi Zhi Institute Innovation Program (No. SQZ202318) and the CPS-Yangtze Delta Region Industrial Innovation Center of Quantum and Information Technology-MindSpore Quantum Open Fund.

\bibliography{ref}

\end{document}